\documentclass[useAMS,usenatbib,referee]{biom}

\usepackage{pdfpages}
\usepackage{lipsum}
\usepackage[T1]{fontenc}
\usepackage[bottom]{footmisc}
\usepackage{amsmath,amsfonts,bbm, dsfont, amsopn}
\usepackage{subcaption}
\usepackage{epsfig}
\usepackage{hyperref}
\usepackage{multicol}
\usepackage[]{caption}
\usepackage{pifont}
\usepackage{enumerate}
\usepackage{epic}
\usepackage{graphicx}
\usepackage{color}
\usepackage{mathtools}
\usepackage{siunitx}
\usepackage{tabularx}
\usepackage{booktabs,collcell,eqparbox}
\usepackage[autostyle]{csquotes}
\usepackage{adjustbox}
\usepackage[english]{babel}
\def\bSig\mathbf{\Sigma}

\newcommand{\R}{\mathbb{R}}
\DeclareMathOperator*{\argmax}{argmax}

\usepackage[figuresright]{rotating}

\title[Probabilistic Canonical Correlation Analysis for Sparse Count Data]{Probabilistic Canonical Correlation Analysis for Sparse Count Data}

\author {Lin Qiu$^{1}$, Vernon M.Chinchilli$^{1}$  \\
	$^{1}$Department of Public Health Sciences and Statistics \\ The Pennsylvania State University, Hershey, PA, U.S.A \\
	{$^{*}$Contact Email: vchinchi@psu.edu}}

\begin{document}

\sloppy








\label{firstpage}


\begin{abstract}
   
Canonical correlation analysis (CCA) is a classical and important multivariate technique for exploring the relationship between two sets of continuous variables. CCA has applications in many fields, such as genomics and neuroimaging. It can extract meaningful features as well as use these features for subsequent analysis. Although some sparse CCA methods have been developed to deal with high-dimensional problems, they are designed specifically for continuous data and do not consider the integer-valued data from next-generation sequencing platforms that exhibit very low counts for some important features. We propose a model-based probabilistic approach for correlation and canonical correlation estimation for two sparse count data sets (PSCCA). PSCCA demonstrates that correlations and canonical correlations estimated at the natural parameter level are more appropriate than traditional estimation methods applied to the raw data. 
We demonstrate through simulation studies that PSCCA outperforms other standard correlation approaches and sparse CCA approaches in estimating the true correlations and canonical correlations at the natural parameter level. We further apply the PSCCA method to study the association of miRNA and mRNA expression data sets from a squamous cell lung cancer study, finding that PSCCA can uncover a large number of strongly correlated pairs than standard correlation and other sparse CCA approaches.
\end{abstract}

%



\maketitle

%

\section{Introduction}
\label{s:intro}

Recent advancements in next-generation sequencing technology have enabled the measurement of multiple high-dimensional data types in a single study, such as genomics, transcriptomics, epigenomics, and metabolomics. Integrative analysis of high-dimensional omics data is becoming increasingly important and popular. It has been shown that combining multiple omics data types can improve analysis and lead to biologically more meaningful results for complex diseases \citep{safo18,Lu19}.

Omics datasets have three main characteristics that pose modeling challenges: (1) they are high-dimensional data, with a large number of variables $p$ and small sample size $n$; (2) the raw data represent count variables which violate the distributional assumptions for standard correlation and canonical correlation analysis, which can lead to invalid inference in the presence of a small sample size; (3) the data are very sparse, with a large proportion of these counts being very close to zero and having random missing values. 

\subsection{\bf{CCA}}

Canonical correlation analysis (CCA) is a classical multivariate method proposed by \citet{Hotelling36} for exploring the relationship between two sets of variables. Consider two random vectors \textbf{X}$\in \R ^{p}$, \textbf{Y}$\in \R ^{q}$. Define \textbf{$\bold{\sum_{XX}}$} = cov(\textbf{X}), \textbf{$\bold{\sum_{YY}}$} = cov(\textbf{Y}), and \textbf{$\bold{\sum_{XY}}$} = cov(\textbf{X,Y}). CCA finds canonical correlation directions  ($\bold{\theta}$,$\bold{\eta}$) that maximize the correlation between \textbf{$a^TX$} and \textbf{$b^TY$}, where the ($\bold{\theta}$,$\bold{\eta}$) are the linear combinations of \textbf{X} and \textbf{Y} canonical variables. Formally, we can write the first pair of canonical variables as

\quad\quad\quad\quad\quad\quad ($\theta,\eta$) = $\argmax_{(\textbf{a},\textbf{b})}$ \{\textbf{$\bold{a^T}$} \textbf{$\bold{\sum_{XY}}$}\textbf{b}: \textbf{$\bold {a^T}$}\textbf{$\bold{\sum_{XX}}$} \textbf{a}=1, \textbf{$\bold{b^T}$}\textbf{$\bold{\sum_{YY}}$} \textbf{b}=1\},

Then the optimization can be attained by applying the singular value decomposition (SVD) and replacing $\bold{\sum_{XX}^{-1/2}} \bold{\sum_{XY}} \bold{\sum_{YY}^{-1/2}}$ with their sample estimates $\bold{\hat{\sum}_{XX}^{-1/2}} \bold{\hat{\sum}_{XY}} \bold{\hat{\sum}_{YY}^{-1/2}}$. However, in a high-dimensional setting, when the dimensions $p, q \gg$ n, the SVD approach is not applicable because $\bold{\hat{\sum}_{XX}}$ and $\bold{\hat{\sum}_{YY}}$ are not invertible.

\subsection{\bf{Related work}}

Motivated by genomics, neuroimaging and other applications, researchers have been working on generalizing CCA to accommodate high dimensions, usually called sparse CCA \citep{Witten09,Avants10,Hardoon11,Gao17}. 
These methods impose sparsity constraints on the canonical directions which effectively can reduce the dimensionality and improve the interpretation of the correlations. Penalized matrix decomposition (PMD) \citep{Witten09}is one of the most popular sparse CCA methods, which uses the penalized matrix decomposition to replace $\bold{\hat{\sum}_{XX}}$ and $\bold{\hat{\sum}_{YY}}$ with identity matrices to avoid singularities. By doing so, PMD can obtain sparse estimates of the canonical directions by penalization. However, PMD may perform poorly on data sets when $\bold{{\sum}_{XX}}$ and $\bold{{\sum}_{YY}}$ are far from diagonal. With respect to genomics data, for example, genes usually have strong correlations among them.

The probabilistic interpretation of CCA was initiated by \citep{Bach06}. 
Later on, several Bayesian versions of CCA were developed \citep{Archambeau08,Virtanen11,Klami13}.  One of the key promising features of Bayesian CCA is that it enables analysis of high-dimensional data in life sciences \citep{Fujiwara09,Huopaniemi10}. However, these methods assume the data to follow normal distributions. Thus, the aforementioned Bayesian methods may not work well for non-normally distributed data. PCAN is the first approach that describes a Bayesian correlation analysis method for count data \citep{Zoh16}, in which the correlations are estimated based on the latent weights from the natural parameters of the data generating model, rather than the correlations among the counts. In the latent variable model, priors or strong regularizations are used on the latent weights to induce sparsity \cite{West03}. 

The rise of big datasets with few signals, such as omics datasets, has spurred the study of sparse models. From a Bayesian perspective, discrete mixtures \citep{George93} and shrinkage priors \citep{Tipping01} are the two main sparse estimation methods. In latent variable models, Bayesian shrinkage priors are popular because of their flexible and interpretable solutions \citep{Carvalho08, Knowles11,Bhattacharya11}. The spike-and-slab prior is a mixture of a point mass at zero and a flat distribution across the space of real values, in that the excluded loadings are modeled by the "spike" distribution, while the included loadings are modeled by the "slab" distribution \citep{Carvalho08}. The disadvantages are that the results can be sensitive to prior choices and it is computationally demanding for posterior inference with a large number of variables due to a huge model space. Scale mixtures of normal priors have been proposed recently as a computationally efficient alternative to the two component spike-and-slab prior \citep{Armagan13, Bhattacharya14}. These types of priors usually assume normal distributions with a mixed variance term and the mixing variance distribution enables strong shrinkage close to zero. For example, Bayesian canonical correlation analysis (BCCA) \citep{Klami13}  consists of applying an automatic relevance determination (ARD) \citep{Neal96} prior for the latent weights which is a Normal-gamma prior that imposes an inverse gamma distribution on the variance term. The horseshoe prior is popular due to its good performance in simulations and under theoretical study, which has shown comparable performance to the spike-and-slab prior in a variety of problems where a sparse prior is desirable \citep{Carvalho08,Carvalho10,Polson11}.%
 The horseshoe prior is a scale mixture of normals, with a product of half-Cauchy priors on the variance. It is given by
\begin{gather*}
\theta_i|\lambda_i,\tau \sim N(0,\lambda_i^2\tau^2),\\
\lambda_i \sim C^+(0,1), i=1,...,n
\end{gather*}

The \enquote{global hyperparameter} $\tau$ can shrink all the parameters toward zero, especially if its domain is restricted to a finite interval, while the heavy-tailed half-Cauchy local priors allow some parameters to escape. Different levels of sparsity can be accommodated by changing the value of $\tau$: the large $\tau$ will have little shrinkage, while small $\tau$ will shrink all the weights to zero. Despite the good performance, there are two shortcomings for the horseshoe prior. First, how to perform inference for the \enquote{global hyperparameter} $\tau$ which determines the overall sparsity in the parameter vector $\boldsymbol{\theta}$ is not fully answered yet. Second, parameters far from zero will not be regularized at all. Quite a few researchers have investigated the impact of $\tau$ concerning the resulting posterior distribution both for recovery and for uncertainty quantification, either in a deterministic way or a hierarchical full Bayes approach \citep{Carvalho08,Datta14,Pas14,Pas17}. We take the second shortcoming as the key strength of this prior and to incorporate it with latent variable model to infer the feature sparsity jointly. For an omics data set we assume only important variables are strongly identified and the parameters far from zero will not be regularized.

\subsection{\bf{Our contribution}}

In this study, we propose a new probabilistic framework of CCA for sparse count data, which we label a probabilistic sparse canonical correlation analysis (PSCCA). Our work contributes several important advances. First, we propose to estimate the canonical correlations at the natural parameter level for data expressed as raw counts, which is lacking in omics analyses. Second, we provide a theoretical justification for estimating the correlations and canonical correlations based on the natural parameters rather than based on the raw data. The former are larger in magnitude than the latter, which is very meaningful for CCA. Because CCA is an exploratory analytical method, larger values of the canonical correlations yield less chance to miss the true correlation pairs. Third, the horseshoe prior is widely studied in the literature, via both simulation studies and theoretical research. Nevertheless, we do not see many examples in applications. We formulate the natural parameters as a latent variable model, and we invoke the horseshoe prior for the latent weight to model the sparsity. To better extract the sparse signals we assume $\tau \sim C^+(0,1)$ for the \enquote{global hyperparameter}. As discussed in \citet{Piironen17}, this prior results in sensible inference only when $\tau$ is strongly identified by the data. Our simulation study and real data applications show that our approach performs better than existing methods. Lastly, our approach is built on an exponential family and can be easily extended to other formats of data.

The rest of the article is organized as follows. Section 2 contains our model details and inference. Section 3 discusses the theoretical results and Section 4 describes simulation studies. Section 5 presents the real data application. Finally, Section 6 contains a discussion and future directions of PSCCA.

\section{Method}
\label{s:model}

\subsection{\bf{Model}}

Let $\textbf{F}_y(\cdot|\cdot)$ be a distribution function from the natural parameter exponential family. The random component of a generalized linear model consists of a response vector $\textbf{y} \in \R^N$ which has a conditional distribution in the exponential family. This family has probability density function or mass function of the form
$f_y(y_j,\theta_j)=a(\theta_j)b(y_j)\text{exp}[y_j\mathcal{Q}(\theta_j)]$. The value of the parameter $\theta_j$ may vary for $j=1,...,N$ depending on values of the explanatory variables. The term $\mathcal{Q}(\theta)$ is the natural parameter; $a(\cdot)$ and $b(\cdot)$ are non-negative functions that distinguish one member of the exponential family from another. For our case,  
assume we have two sets of multivariate random variables, $\boldsymbol{Y}^{(1)} \in \R^{D_1\times 1}$,
and $\boldsymbol{Y}^{(2)} \in \R^{D_2\times 1}$. 
The observed data samples are expressed as $[\boldsymbol{Y}_1^{(m)},..., \boldsymbol{Y}_N^{(m)}]\in \R^{D_m\times N}$ with $N$ observations, where $m$ is 1 or 2.
Let $y_{ij}^{(m)}$ represent the observed value of the $j^{th}$ individual for the $i^{th}$ feature (variable) in a set of $D_m$ measured features (variables).

We motivate our formulation in the latent variable interpretation of CCA \citep{Bach06} to model the natural parameters and the ideas from PCAN (Zoh et al., 2016) for the correlation estimation from the natural parameters. We assume each individual data vector follows conditionally an exponential family distribution and here we consider a generalized linear model. The generative model for $D_m$ coupled natural parameters $\boldsymbol{\theta}_{.j}^{(m)}$ with $m$=1,2 and $j=1,...,N$ is

\begin{equation}
\begin{aligned}
\label{eq1}
&{\boldsymbol\theta}_{.j}^{(m)}={\boldsymbol \mu}_{\theta}^{(m)} + {\bf W}^{(m)} {\bf Z}_j + {\boldsymbol \epsilon}_{.j}^{(m)}, 
\end{aligned}
\end{equation}

We write the model as a function of latent variables by concatenating the $D_m$ features into the vector ${\boldsymbol \theta}_{.j}^{(m)}$ 

\begin{equation} 
\begin{aligned}
&{\boldsymbol Y}_{.j}^{(m)}|{\boldsymbol \theta}_{.j}^{(m)}\sim \text{Poisson} \big \{ \text {exp}({\boldsymbol \theta}_{.j}^{(m)}) \big \},\\
&{\boldsymbol \epsilon}_{.j}^{(m)} \sim { f}_\epsilon({\boldsymbol\epsilon}_{.j}^{(m)}),\\
&{\bf Z}_j \sim N_d(0,{\bf I_{d\times d}}).
\end{aligned}
\end{equation}

The matrix $\textbf{W}^{(m)}\in \R^{D_m\times d}$ denotes the loading matrix associated with the latent vector ${\bf Z}_j=(Z_{1j},...,Z_{dj})^T $; ${\bf I_d}$ denotes the $d\times d$ identity matrix; the parameter vector ${\boldsymbol \mu}_{\theta}^{(m)}$ represents the mean of the natural parameters associated with the $i^{th}$ feature in the data sets ${\boldsymbol Y}_{.j}^{(m)}$; and ${\boldsymbol \epsilon}_{.j}^{(m)}$ is an independently distributed error term, with $f_\epsilon$ denoting a normal distribution with null mean and variance $\sigma_{\epsilon}^2$. The core generative process is the unobserved shared latent variables ${\bf Z}_j$, which are transformed via linear mappings to the observation spaces, and can capture the variation common to both data sets and allow for dependency between variables in a specific data set.

We impose horseshoe priors on $D_m \times d$ matrix $\textbf{W}^{(m)}$, let ${\boldsymbol W}_{i.}^{(m)}$ denote the ith row vector of ${\boldsymbol W}^{(m)}$. Then we assume that:

\begin{equation}
\begin{aligned}
& {\boldsymbol W}_{i.}^{(m)}|\lambda_i^{(m)},\tau^{(m)} \sim \text{N}({\boldsymbol W}_{i.}^{(m)}|{\boldsymbol 0},{\lambda_i^{(m)}}^2 {\tau^{(m)}}^2{\bf I}), 
\end{aligned}
\end{equation}
We refer to the $\lambda_i^{(m)}$ as the local shrinkage parameters and to $\tau^{(m)}$ as the global shrinkage parameters. Let $C^+(0,1)$ denote the half-Cauchy distribution. The half-Cauchy prior for the local shrinkage parameter $\lambda_i$ has shown good performance \citep{Carvalho08,Carvalho10}. There has been a vast amount of research on how to choose the prior for the global hyperparameter $\tau$ which plays an important role in overall sparsity in the parameter matrix ${\boldsymbol W}^{(m)}$. As discussed in Introduction section, we choose the full Bayesian specification for $\tau$. Thus, we assume:

\begin{equation}
\begin{aligned}
& \lambda_i^{(m)}\sim C^+(0,1); \tau^{(m)} \sim C^+(0,1)\\
\end{aligned}
\end{equation}

For i =1,...,$D_1$; k=1,...,$D_2$; and j=1,...,N, we construct $\boldsymbol{\theta}_{.j}^{(1)} =({\theta}_{1j}^{(1)},...,{\theta}_{D_1j}^{(1)})^T$ and $\boldsymbol{\theta}_{.j}^{(2)} =({ \theta}_{1j}^{(2)},...,{\theta}_{D_2j}^{(m)})^T$. The vector $(\boldsymbol{\theta}_{.j}^{(1)} \quad \boldsymbol{\theta}_{.j}^{(2)})^T$ has a multivariate normal distribution with mean \\
${\boldsymbol{ \mu}_{\theta}}=(\mu_{\theta_1}^{(1)},...,\mu_{\theta_{D_1}}^{(1)},\mu_{\theta_1}^{(2)},...,\mu_{\theta_{D_2}}^{(2)})^T$ and covariance matrix 

\begin{equation}
\boldsymbol{\Sigma}= \left(\begin{array}{cc}
{\bf W}^{(1)}{{\bf W}^{(1)}}^T + {\sigma_\theta^{(1)}}^2{\bf I_1} & {\bf  W}^{(1)}{{\bf W}^{(2)}}^T \\
{\bf W}^{(2)} {{\bf W}^{(1)}}^T & {\bf W}^{(2)}{{\bf W}^{(2)}}^T + {\sigma_\theta^{(2)}}^2{\bf I_2}	
\end{array}\right)
\end{equation}
where ${\bf I_1}= {\bf I_{D_1\times D_1}}$ and ${\bf I_2}= {\bf I_{D_2\times D_2}}$.
\vspace{0.5\baselineskip}  
The correlation between $\theta_{ij}^{(1)}$ and $\theta_{kj}^{(2)}$, for any sample $j$ can be obtained as 

\begin{equation}
\begin{aligned}
&\text{corr}(\boldsymbol{\theta}_{.j}^{(1)},\boldsymbol{\theta}_{.j}^{(2)})= ({\bf W}^{(1)}{{\bf W}^{(1)}}^T  + {\sigma_\theta^{(1)}}^2{\bf I_1})^{-1/2}{\bf  W}^{(1)}{{\bf W}^{(2)}}^T
({\bf W}^{(2)}{{\bf W}^{(2)}}^T+ {\sigma_\theta^{(2)}}^2{\bf I_2})^{-1/2}
\end{aligned}
\end{equation}
\vspace{0.5\baselineskip}

For the canonical correlations, let ${\boldsymbol C}^{-1/2}$ denote the square-root decomposition of the positive definite matrix $ {\boldsymbol C}^{-1}$. Let $\textbf{R}=({\bf W}^{(1)}{{\bf W}^{(1)}}^T + {\sigma_\theta^{(1)}}^2{\bf I_1})^{-1}{{\bf  W}^{(1)}{{\bf W}^{(2)}}^T}({\bf  W}^{(1)}{{\bf W}^{(2)}}^T + {\sigma_\theta^{(2)}}^2{\bf I_2})^{-1}{\bf W}^{(2)} {{\bf W}^{(1)}}^T $, then the nonnull eigenvalues of \textbf{R} correspond to the squared canonical coefficients for the natural parameters. Inference on the correlations and canonical correlations will be based on the marginal posterior distribution of ${\bf W}^{(1)}$, ${\bf W}^{(2)}$, ${\sigma_\theta^{(1)}}^2$, and ${\sigma_\theta^{(2)}}^2$.

\subsection{\bf{Identifiability and Prior}}

The latent variable model (Equation 1) is identifiable up to orthonormal rotations, for any invertible $\textbf{G}\in \mathcal{R}^{d\times d}$ with  $\textbf{G}^T\textbf{G} = \textbf{I}$. Then  $\textbf{W}^* = \textbf{W}\textbf{G}^T $ and $\textbf{Z}^*=\textbf{G}\textbf{Z}$ will produce the same estimate of the data covariance matrix in Equation (5) and has an equal likelihood. \citet{Zoh16} recommend imposing a lower triangular structure for the matrices ${\bf W}^{(m)}$, following the work of \citep{Geweke96}, and further require that the diagonal elements are non-negative to remove the non-identifiability related to the sign. This approach relies on the choice of $d$ based on the $D_m$ dimensions of \textbf{W}, but we assume the value of $d$ will be small so that the impact is negligible \cite{Lopes04}.

\begin{equation}
\begin{aligned}
& W_{ik}^{(m)}\sim \text{N}(W_{ik}^{(m)}|0,{\lambda_i^{(m)}}^2 {\tau^{(m)}}^2) if i < k,\\
& W_{ik}^{(m)}\sim \text{N}(W_{kk}^{
(m)}|0,{\lambda_i^{(m)}}^2 {\tau^{(m)}}^2){\bf 1}(W_{kk}^{(m)}>0) if i = k,
\end{aligned}
\end{equation}

where for $m=1$, $i=1,...,D_1, k=1,...,d$ and for $m=2$, $i=1,...,D_2, k=1,...,d$. \textbf{1} is an indicator function, \textbf{1(W)}=\textbf{1} if W is true and 0 otherwise. We assume conjugate priors for the remaining parameters in the model as 
\begin{equation}
\begin{aligned}
&{\boldsymbol \mu}_{\theta}^{(m)}\sim  \Pi_{i=1}^{D_m}\text{Normal}(0,k_{i}^{(m)}),\\
&{\sigma_{\theta}^{(m)}}^2\sim \text{Inv-}\chi^2(\nu_{\theta}^{(m)},{s_{\theta}^{(m)}}^2), 
\end{aligned}
\end{equation}
The hyperparameters  ${ k}_i^{(m)}$, ${s_{\theta}^{(m)}}^2$, and $\nu_{\theta}^{(m)}$ are known.

\subsection{\bf {Inference}}
\label{s:inf}

The form of the full conditional posterior distribution is proportional to the product of (1) the joint conditional likelihood for the data matrices ${\bf Y}^{(1)}$ and ${\bf Y}^{(2)}$ and (2) the prior distributions:
\begin{equation}
\begin{aligned}
&P(\boldsymbol{\theta}^{(1)} ,\boldsymbol{\theta}^{(2)},{\bf W}^{(1)},{\bf W}^{(2)},{\bf Z},{\sigma_{\theta}^{(1)}}^2,{\sigma_{\theta}^{(2)}}^2|{\bf Y}^{(1)},{\bf Y}^{(2)})\\
&\propto l({\bf Y}^{(1)},{\bf Y}^{(2)}|\boldsymbol{\theta}^{(1)} ,\boldsymbol{\theta}^{(2)},{\bf W}^{(1)},{\bf W}^{(2)},{\bf Z},{\sigma_{\theta}^{(1)}}^2,{\sigma_{\theta}^{(2)}}^2)\\
&\times \Pi_{i=1}^d \Big \{ (\lambda_i^{(1)} \tau^{(1)})^{D_1}\text{exp}(-0.5\lambda_i^{(1)} \tau^{(1)} {\bf W}_{i.}^{(1)}{{\bf W}_{i.}^{(1)}}^T)\\
&(\lambda_i^{(2)} \tau^{(2)})^{D_2}\text{exp}(-0.5\lambda_i^{(2)} \tau^{(2)} {\bf W}_{i.}^{(2)}{{\bf W}_{i.}^{(2)}}^T)\\
& \times \frac{1}{1+{\lambda_i^{(1)}}^2}\frac{1}{1+{\tau^{(1)}}^2}\\
& \times \frac{1}{1+{\lambda_i^{(2)}}^2}\frac{1}{1+{\tau^{(2)}}^2} \Big \} \\
& \times \Big \{\Pi_{j=1}^N\text{exp}(-0.5 {\bf Z_j^TZ_j})\Big \}\text{exp}\{-\nu_{\theta}^{(1)}{s_{\theta}^{(1)}}^2/(2{\sigma_{\theta}^{(1)}}^2)\}\\
&\times{\sigma_{\theta}^{(1)}}^{-2(1+\nu_{\theta}^{(1)}/2)}\\
& \times \text{exp}\{-\nu_{\theta}^{(2)}{s_{\theta}^{(2)}}^2/(2{\sigma_{\theta}^{(2)}}^2)\}{\sigma_{\theta}^{(2)}}^{-2(1+\nu_{\theta}^{(2)}/2)}.
\end{aligned}
\end{equation}

The posterior distribution in (9) is difficult to directly simulate. We update the parameters in a Markov chain Monte Carlo (MCMC).

\section{Theoretical Results}
 Let $\textbf{F}_x(\cdot|\cdot)$ and $\textbf{F}_y(\cdot|\cdot)$ be distribution functions from the natural parameter exponential family as we discussed in the model section. We model $X_i|\theta_i \sim  \textbf{F}_x(\theta_i)$ and $Y_k|\lambda_k \sim  \textbf{F}_y(\lambda_k)$, and we denote ${ X}_{i}|{ \theta}_i\sim \text Poisson \big \{ \text {exp}({\theta}_i) \big \},i=1,2,...,p$ and ${Y}_{k}|{ \lambda}_k\sim \text Poisson \big \{ \text {exp}({\lambda}_k) \big \},k=1,2,...,q$, $p < q$, as two sets of count random variables. The natural parameters $\theta_i$ and $\lambda_k$ have the following expressions:
 
 \begin{equation}
 \begin{aligned}
 {\theta}_i=\mu_{\theta i} + {\bf W}_{i.}^1 {\bf Z} +  \epsilon_i and {\lambda}_k=\mu_{\lambda k} + {\bf W}_{k.}^2 {\bf Z} +  \eta_k
 \end{aligned}
 \end{equation}
 where ${\bf W}^1$ is $p \times d$ and ${\bf W}^2$ is $q \times d$. We further assume the following independent probability distributions:
  ${\bf Z} \sim N_d({\bf 0}_{d\times 1}$,${\bf I_{d\times d}})$,
  $\epsilon_1,\epsilon_2,...,\epsilon_p i.i.d \sim N(0,\sigma_{\epsilon}^2)$, $\eta_1,\eta_2,...,\eta_q i.i.d \sim N(0,\sigma_{\eta}^2)$.\\

\textbf{Theorem 1.} {\itshape We define the unconditional variance-covariance of  
${\bf X}$ and ${\bf Y}$ as \textbf{$\bold{\sum_{XY}^{**}}$}.
 Then we have the correlation coefficients, $|\varphi_{lm}| \leq \omega$, constructed from \textbf{$\bold{\sum_{XY}^{**}}$}, where $0 <\omega < 1$, $l=1,...,p$, and $m=1,...,q$. The canonical correlation coefficients $\varrho_{lm} \leq \psi$, where $0< \psi < 1$.  The detailed proof is in the Appendix}.\\
 
\textbf{Corollary 1.}
{\itshape Correlation coefficients and canonical correlation coefficients calculated from the raw count data {\bf X} and {\bf Y} will be smaller numerically in magnitude than the correlation coefficients and canonical correlation coefficients calculated from the natural parameters ${\bold{ \theta}}$ and ${\bold {\lambda}}$}.

\section{Simulation}
\subsection{\bf{Settings}}
\label{s:discuss}

In this section we conduct simulations to assess the performance of SPCCA in comparison with several existing methods that have been proposed for probabilistic correlation analysis (PCAN), Bayesian CCA (BCCA), and sparse CCA (PMD), as mentioned in the Introduction. First, we evaluate the performance of SPCCA on correlation analysis, comparing with PCAN because the PCAN paper already demonstrated that it outperforms traditional Spearman and Pearson correlation methods. Second, we compare SPCCA's performance on the canonical correlation analysis with that of BCCA, PMD, and the modified PCAN, which we name $ \text{PCAN}^{*}$. 

To evaluate the methods, let $ \bf{W}^{(1)*} = (w_1^{(1)*},...,w_{D_1}^{(1)*})$, $ \bf{W}^{(2)*} = (w_1^{(2)*},...,w_{D_2}^{(2)*})$ be the true generated loading matrices. For the all methods with estimates $\bf{\hat{W}}^{(1)} = (\hat{w}_1^{(1)},...,\hat{w}_{D_1}^{(1)})$, $ {\bf \hat{W}}^{(2)} = (\hat{w}_1^{(2)},...,\hat{w}_{D_2}^{(2)})$, we can calculate the correlations and the canonical correlations according to Equations (5) and (6). Let $\textbf{U}^{D_1\times D_1}$ be the true matrix and $\textbf{V}^{D_2\times D_2}$ be the estimated matrix. Then we construct the Frobenius loss function as $\sum_{i,j}({\bf U_{ij}}-{\bf V_{ij}})^2$, assuming $D_1 < D_2$.


\textbf{Scenario I: Correlation analysis}-- In the first scenario, we simulated 100 datasets assuming for each dataset $D_1=10$, $D_2=30$, and $N=50$ subjects. The weight matrices are $\textbf{W}_{D_1\times d}^{(1)}$ and $\textbf{W}_{D_2\times d}^{(2)}$. We consider three correlation matrices for the natural parameters: \\
(a) the identity correlation matrix assuming the true $d=0$ for $\textbf{W}_{D_1\times d}^{(1)}$ and $\textbf{W}_{D_2\times d}^{(2)}$.\\
(b) a correlation matrix obtained assuming $d=5$ for $\textbf{W}_{D_1\times d}^{(1)}$ and $\textbf{W}_{D_2\times d}^{(2)}$.\\
(c) a correlation matrix obtained assuming $d=10$ for $\textbf{W}_{D_1\times d}^{(1)}$ and $\textbf{W}_{D_2\times d}^{(2)}$.\\

We fit the PSCCA and PCAN to each of these 100 datasets assuming different dimensions of $d$ to compute the posterior mean correlation matrices. 
\begin{table}[ht]
\begin{center}
\caption{\itshape Summary of the Frobenius loss when estimating the true correlation structure for the natural parameters from the PCAN and our PSCCA model. Here, $d$ is the value of $d$ assumed for the true correlation matrix; $d^*$ represents the value $d$ assumed when fitting the model. Frobenius losses are calculated between the true correlation matrix at the natural parameter level vs the posterior mean correlation estimated based on the posterior of $\textbf{W}^{(1)}$, $\textbf{W}^{(2)}$ and the other parameters.}
\label{s:table1}

\begin{tabular}{ccc ccc}
  \toprule
  \toprule
  
 &&\multicolumn{2}{c}{PSCCA} & 
				\multicolumn{2}{c}{PCAN} \\ \cline{3-4} \cline{5-6}
  $d$ & $d^*$ & {Mean} & {95\%CI}  &{Mean} &  {95\%CI}\\ \midrule
            0& 2& 25.21 & (24.35, 26.54) &		28.57&	(28.31, 29.20)  \\
0&5&  	19.55	& (19.20, 20.49) & 23.30&	(21.85, 24.95)  \\
0&10& 14.12&	(13.23, 14.83) & 21.26&	(19.66, 22.17) \\
\\
5&2& 4.25&	(4.05, 4.34) &   22.51& (21.26, 22.87)  \\
5&5& 3.89& (3.60, 4.24) & 19.88&	(18.72, 20.63) \\
5&10& 4.54&	(4.21, 4.91) & 20.35&	(19.45, 22.44) \\
\\

10&2& 10.31 & (9.41, 12.21) &   15.55&(14.68, 16.31) \\
10&5& 8.81 &(8.26, 9.35) & 14.62 &(13.34, 15.21)	\\
10&10& 7.85 &(7.54, 8.06) & 13.27 & (12.82, 13.51) \\

		  \bottomrule
\end{tabular}
\end{center}
\end{table}

\textbf{Scenario II: Canonical correlation analysis}-- In the second scenario, we simulated 100 datasets, for each dataset we set $N=100$, $d=10$ under moderate and high dimensions of $D^{(m)}$. We use three models for the correlation matrices of the the natural parameters $\boldsymbol{\theta}^{(m)}$.\\
Model I (Independent covariances): there is no covariance structure within each of the natural parameters $\boldsymbol{\theta}^{(m)}$. \\
Model II (Identity covariances): $\sum_{\boldsymbol \theta ^{(1)} \theta^{(1)}} = {\bf I}$,  $\sum_{\boldsymbol \theta ^{(2)} \boldsymbol \theta ^{(2)}} = {\bf I}$\\
Model III (Moderate covariances): $\sum_{\boldsymbol \theta ^{(1)} \boldsymbol \theta ^{(1)}}$ = {\bf 0.5},  $\sum_{\boldsymbol \theta ^{(2)} \theta ^{(2)}}$ = {\bf 0.5}\\
(1) Moderate dimensions: $D_1 = D_2$ = 60, 100, 300\\
(2) High dimensions: $D_1 = D_2$ = 500, 1000, 2000\\

Model I is used in PCAN \citep{Zoh16}, and similar models of Model II and III have been used to generate the raw data in \citep{Gao17}.  We fit PSCCA, $\text{PCAN}^{*}$, PMD, and BCCA to the datasets simulated under the above scenario and compute the canonical correlation matrices for the purpose of comparison.  
\subsection{\bf{Results}}
We report the Frobenius loss in (Table 1) and Stein loss (Supplementary Table 1) for each of the estimated correlation matrices.  Stein loss is defined as $\textbf{diag} ({\bf V^{-1}U})-\textbf{det}({\bf V^{-1}{\bf U}}) - D_1$ for estimating the $D_1\times D_1$ matrix \textbf{V} and the $D_1\times D_1$ matrix \textbf{U}. 

Overall, under each of the scenarios, PSCCA yields a smaller Frobenius loss and a smaller Stein loss compared to PCAN, and both methods yield much smaller Frobenius losses compared to Stein losses. We found that under the true $d$=5, 10, when the assumed $d^*$ is closer to the truth, PSCCA and PCAN result in smaller Frobenius losses, and smaller losses are preferred. However, when $d=0$ we observe the opposite situation in that the closest value to the truth when $d^*$=2 yields the largest Frobenius loss for PSCCA and PCAN. 

We also estimate the correlations using other standard correlation estimation methods. Because we assumed $N > D_1 + D_2$, other standard correlation estimate methods are valid. We report the summary of the Frobenius loss incurred with estimating the true correlation matrices using Pearson and Spearman approaches based on the raw data in Supplementary Table 2. PSCCA resulted in a smaller Frobenius loss, whereas PCAN and Spearman correlations perform similarly under the true $d$ = 0, 10, which is consistent with the results in PCAN paper \citep{Zoh16}. 

For the canonical correlation analysis comparison, we modified the method of PCAN based on Equations (5) and (6) to render it as an alternative approach for a probabilistic canonical correlation analysis method, which we named $\text{PCAN}^{*}$. We report the summary of the results compared with $\text{PCAN}^{*}$, BCCA, and PMD under Models I-III and moderate and high dimensions of $D_m$ in Table 2. 

\begin{table}[htbp]
  \centering
  \newcommand{\pa}[1]{\eqmakebox[CFNetvsNeuMF][r]{#1}}
  \newcommand{\hlc}{\textbf}
  \caption{\itshape Simulation results for Models I-III under moderate and high dimensions of $D_m$. The reported numbers are the medians and standard errors (in parentheses) of canonical correlation's   Frobenius loss over 100 replicates.}
  \label{s:table2}
  \begin{adjustbox}{max width=1.1\textwidth,center}
  \begin{tabular}{ *{10}{c} >{\collectcell\pa}c<{\endcollectcell} }
    \toprule
    \toprule
    &
    \multicolumn{3}{c}{$D_1=D_2=60$} &
    \multicolumn{3}{c}{$D_1=D_2=100$} &
    \multicolumn{3}{c}{$D_1=D_2=300$}
    \\
    \cmidrule(lr){2-4}\cmidrule(lr){5-7}\cmidrule{8-10}
     & 
    Mode I & 
    Mode II & Model III & Model I & Model II &
    Model III & Model I & Model II & Model III\\
    \midrule
    PSCCA &  0.8178  & 1.102 & 0.8706 & 1.142 & 1.287 &
      1.082 & 1.263 & 1.124 &
      1.349 \\
          & (0.1321) & (0.075) & (0.0191) & (0.037) & (0.1083) &
      (0.0652) & (0.014) & (0.0152) &
      (0.0352) \\
    \addlinespace
    PCAN* &  0.1258  & 1.115 & 1.135 & 1.597 & 1.347 &
      1.272 & 1.594 & 1.295 &
      1.726 \\
           & (0.1459) & (0.0649) & (0.0102) & (0.1710) & (0.028) &
      (0.0687) & (0.022) & (0.0159) &
      (0.0913) \\
    \addlinespace
    PMD &  0.1246  & 1.094 & 1.569 & 1.445 & 1.293 &
      1.450 & 1.495 & 1.135 &
      1.742 \\
           & (0.049) & (0.0344) &(0.0342) & (0.033) & (0.0253) &
      (0.4562) & (0.009) & (0.0235) &
      (0.4289) \\
    \addlinespace
    BCCA & 1.436  & 1.181 & 1.035 & 1.328 & 1.362 &
      1.156 & 1.292 & 1.7997 &
    1.543 \\
         & (0.2462) & (0.2739) & (0.2268) & (0.3551) & (0.281) &
      (0.1596) & (0.276) & (0.0197) &
    (0.2039) \\
      \midrule
      
    &
    \multicolumn{3}{c}{$D_1=D_2=500$} &
    \multicolumn{3}{c}{$D_1=D_2=1000$} &
    \multicolumn{3}{c}{$D_1=D_2=2000$}
    \\
    \cmidrule(lr){2-4}\cmidrule(lr){5-7}\cmidrule{8-10}
     & 
    Mode I & 
    Mode II & Model III & Model I & Model II &
    Model III & Model I & Model II & Model III\\
    \midrule  
    PSCCA &  1.124  & 1.307 & 1.381 & 1.448 & 1.265 &
      1.611 & 1.220 & 1.532 &
      1.542 \\
          & (0.0076) & (0.0570) & (0.0158) & (0.0014) & (0.0104) &
      (0.058) & (0.0020) & (0.0368) &
      (0.0267) \\
    \addlinespace
    PCAN* &  1.737  & 1.407 & 1.547 & 1.606 & 1.842 &
      1.661 & 1.609 & 1.842 &
      1.589 \\
           & (0.0138) & (0.0323) & (0.0152) & (0.0043) & (0.0302) &
      (0.0934) & (0.0017) & (0.0302) &
      (0.0640) \\
    \addlinespace
    PMD &  1.471  & 1.375 & 1.681 & 1.518 & 1.684 &
      1.694 & 1.263 & 1.591 &
      1.783 \\
           & (0.0037) & (0.0121) &(0.5264) & (0.0043) & (0.5903) &
      (0.4157) & (0.0050) & (0.4013) &
      (0.4106) \\
    \addlinespace
    BCCA & 1.332  & 1.497 & 1.589 & 1.603 & 1.717 &
      1.788 & 1.391 & 1.669 &
    1.969 \\
         & (0.4000) & (0.4430) & (0.2452) & (0.3021) & (0.1845) &
      (0.2262) & (0.3900) & (0.2331) &
    (0.3903) \\

    \bottomrule
  \end{tabular}
  \end{adjustbox}
\end{table}

\textbf{Moderate dimensions}.
PSCCA uniformly outperforms the three competitors. The estimator of PSCCA is closer to the truth than the estimates given by $\text{PCAN}^{*}$, PMD, and BCCA. It also is worth noting that under Model II, PMD performs similarly with PSCCA because we generated the natural parameters with identity variance matrices. However, under Model I and Model III, PMD performs poorly. This confirms that methods with no assumptions on the variance matrices have broader applicability. Another point worth noting is that BCCA produces the largest standard errors compared to other methods, which indicates very unstable estimation for count data.

\textbf{High dimensions}. PSCCA continues to outperform the competitors when the dimensions are very high. PMD still displays similar performance with PSCCA under Model II. This suggests when the identity variance assumption holds, the performance of PMD can be improved. when the dimension exceeds 1000 under Model II, however, PMD displays very large standard errors. Under Model III, $\text{PCAN}^{*}$ performs nearly as well as PSCCA, which indicates there exists a moderate level of correlation and the canonical correlation estimated from the natural parameters is closer to the truth. BCCA still has large standard errors compared to other methods, which indicates it is not a proper method for count data.

\section{\bf{Real Data Analysis}}
The Cancer Genome Atlas (TCGA) (http://cancergenome.nih.gov) was initiated in 2006 to develop a publicly accessible infrastructure data on an increasing number of well-characterized cancer genomes. TCGA finalized tissue collection with matched tumor and normal tissues from 11,000 patients with 33 cancer types and subtypes, including 10 rare types of cancer. TCGA data has been used to characterize key genomic changes, find novel mutations, define intrinsic tumor types, discover similarities and differences across cancer types, reveal therapy resistance mechanisms, and collect tumor evolution evidence \citep{Tomczak15}.

MicroRNAs (miRNA) are very short non-coding RNAs that regulate gene expression at the post-transcriptional level. They bind to mRNAs and inhibit translation or induce mRNA degradation. There are many studies that demonstrate negative correlations in the expression of specific miRNAs and their corresponding target mRNAs, and their interaction in many disease-related regulatory pathways is well established \citep{Ruike08,Wang09,Shah11}. In recent years, there have been numerous studies about miRNA and mRNA correlation analysis on different cancer diseases using TCGA data \citep{Ding19,Yu19}. However, these correlation analyses all are based on the normalized continuous data, not the raw count data. 

In our analysis, we consider the read count next generation sequencing (NGS) expression data from squamous cell lung cancer (LUSC). We downloaded the LUSC dataset from TCGA data portal. LUSC has 504 samples, and we processed the tumor miRNA and mRNA data according to TCGAbiolinks \citep{Colaprico16}. Each sample contains 1,881 miRNAs and 56,537 mRNAs. Firstly, we are interested in the correlation analysis between low-expressed mRNAs and a given set of miRNAs. We consider $N=100$ matched miRNA and mRNA samples, and we select $D_1=50$ miRNAs and $D_2=60$ mRNAs. For miRNAs we choose some reported with high correlations with mRNAs in PCAN \citep{Zoh16}, and among 60 mRNAs we choose 30 mRNAs with average counts between 1 and 2, and the remaining 30 mRNAs are the significant expressed genes reported by \citep{Shah11}.
\begin{figure}[ht]
\begin{center}
	\includegraphics[width=6in]{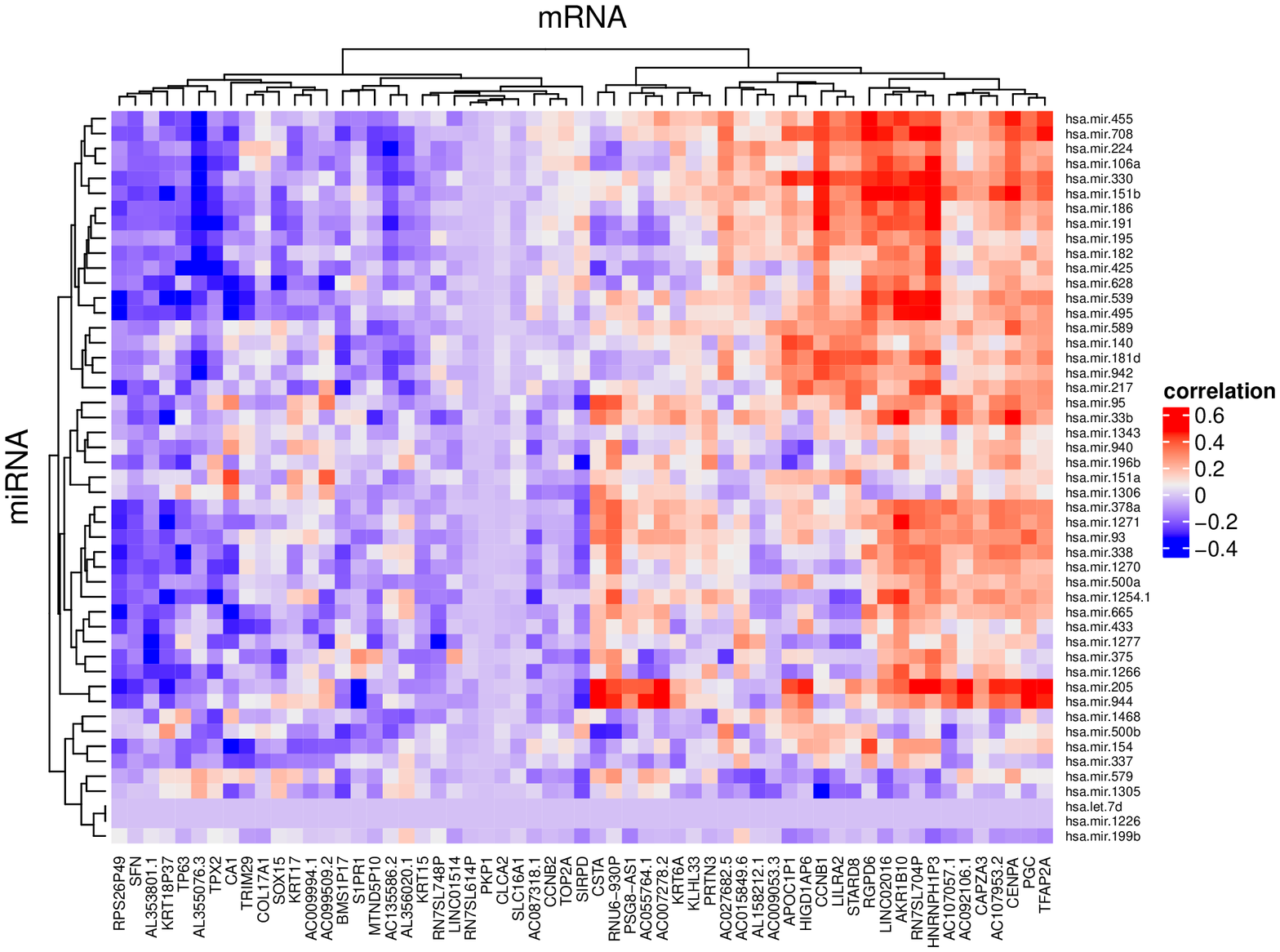}
	\caption{\itshape PSCCA heatmap of the posterior mean correlation estimates between the miRNA and mRNA under $d=10$. Red color indicates the positive correlation and blue color indicates the negative correlation.}
	\label{s:figure1}
\end{center}
\end{figure}
We fit the PSCCA model using the priors in Appendix A under $d=2,5,10$. We ran two separate MCMC chains for 10,000 iterations, and monitored them for proper mixing. The first 5,000 iterations were discarded as burn-in and the inference was based on the 10,000 remaining iterations. We estimate the correlations between miRNA and mRNA based on the posterior mean values of natural parameters, and we also report results based on the standard correlation estimation approaches (Spearman and Pearson) applied to the raw data for comparison. We display, as a heatmap, the posterior mean estimates of each correlation in Figure 1. 

The correlation results identify very interesting miRNA-mRNA interactions. PSCCA demonstrated the highest power to select the potentially correct miRNA-mRNA interactions. Our results show that miR-539 is negatively correlated with genes $RPS26P49$, $KRT18P37$, $TP63$, and $CA1$. The gene encoding miR-539 is located on human chromosome 4q32.31, and miR-539 has been reported to be down-regulated in many human cancers, including prostate cancer, nasopharyngeal carcinoma and thyroid cancer, and  miR-539 has been reported to play a tumor suppression role in many human malignancies \citep{Guo18}. Through TargetScanHuman we confirmed that $TP63$ is the target gene of miR-539, and all the methods estimated the correct negative correlation direction. However, PSCCA has the lowest estimated correlation value -0.3102 between miR-539 and $TP63$ compared to -0.1054 in PCAN, -0.1149 in Spearman, and -0.1218 in Pearson. Meanwhile, from TargetScanHuman, we noticed that miR-539 also regulates $KRT13$, $CA11$ which are the same protein family of $KRT18P37$, and $CA1$, respectively. Thus, our findings might add new members for the target gene family of miR-539, and provide more clues for miR-539's regulation role in lung cancer. Another interesting miRNA is miR-205 which was reported to play a dual role, depending on the specific tumor type and target genes \citep{Nordby17}, we found it to be negatively correlated with $S1PR1$, $RPS26P49$, $SFN$, and $SLC16A1$ in our study.  $S1PR1$ is the target validated through TargetScanHuman, and the estimated correlation value between $S1PR1$ and miR-205 in PSCCA is -0.2875 compared to -0.0896 in PCAN, -0.00617 in Spearman, and -0.01743 in Pearson. The same situation occurred for the mir-338 and TP63 interaction in which all the methods estimated the correct negative correlation directions, but PSCCA has the most extreme negative value compared to other three methods. Here, we report a few interesting miRNA and their estimated correlation with mRNAs in Figure 2(a). From Figure 2(a), we can see that for the same pair of miRNA-mRNA that our PSCCA can estimate the most extreme correlation values among all the other methods.
\begin{figure}[ht]
\begin{center}
	\includegraphics[width=5in]{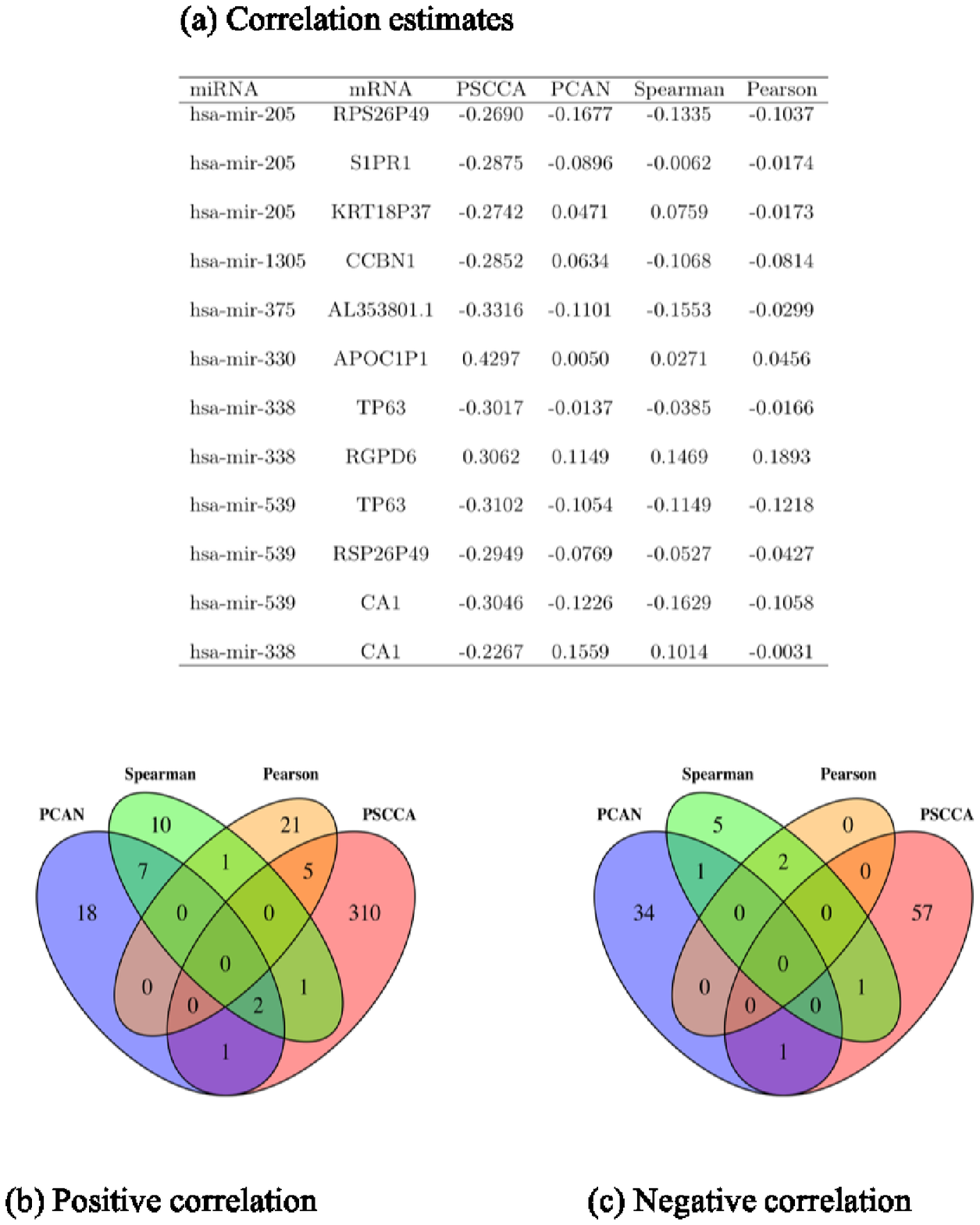}
	\caption{\itshape Correlation estimates (a) and Venn diagram summarising the positive correlation greater than 0.25 (b) and negative correlation less than -0.25 (c) under PSCCA, PCAN, Spearman, and Pearson. }
	\label{s:figure2}
\end{center}
\end{figure}
Figure 2(b) and Figure 2(c) show the venndiagram depicting the overlap between the pairs miRNA-mRNA correlation idenfied by PSCCA, PCAN, Spearman and Pearson approaches. Overall, there is very few pairs jointly identified by four approaches, PSCCA identified 59 negative correlation pairs with estimated correlation values less than -0.25, compared to 36 in PCAN, 9 in Spearman, and 2 in Pearson. For positive correlations, PSCCA identified 319 pairs with estimated correlation values larger than 0.25, compared to 28 in PCAN, 27 in Pearson, and 21 in Spearman. These results suggest PSCCA can identify more extreme negative and positive correlation estimates in sparse count data, and thus have less chance to miss the true correlation pairs for high-dimensional exploratory analysis. For detailed results please check the Supplementary material. 

For canonical correlation analysis, we modified PCAN to calculate the canonical correlation from the natural parameters, termed $ \text{PCAN}^{*}$. We apply PSCCA, $\text{PCAN}^{*}$, PMD,  and BCCA on the same LUSC dataset as above for sample size $N=100$, $D_1=50$ miRNAs, and $D_2=60$ mRNAs. Here, for ease of presentation, we focus only on the first two canonical correlations. We fit all the models on $d=2,5,10$, that is, the number of canonical vectors to be obtained. For PMD, we use equal tuning parameters $\lambda_{a_1}=\lambda_{\beta_1}$, in which the tuning parameter is chosen by the function CCA.permute in the R package PMA. For BCCA, we use the default settings for initial parameter values. PSCCA and ${\text{PCAN}^*}$ yield high canonical correlations, while PMD and BCCA do not perform very well, both yielding small canonical correlations. In addition, under $d=2$ $\text{PCAN}^{*}$ has slightly larger canonical correlation values than PSCCA, which might be because the small $d$ cannot capture the variance about the truth. To check the reasons for poor performance of PMD and BCCA, we found that the variance matrices of the two data sets {\bf X} and {\bf Y} are distant from identity matrices, which severely violates the identity variance assumption imposed by PMD. Also, the data are very sparse because we selected 30 out of 60 mRNAs with the average count between 1 and 2 and the remaining 30 mRNAs have large count values which severely violates the standard normal distribution assumption on the two data sets made by BCCA. PMD and BCCA still can run on this low-dimensional dataset; however, when we apply those methods on the miRNA and mRNA data sets with high dimensions ($D_1=D_2=1000)$, PMD was shut down directly, while BCCA yields the first canonical correlation value of 0.4174 which is far less than PSCCA 0.95 under $d=10$. That again indicates that for high-dimensional sparse count data in genomics, efficient and more accurate canonical correlation methods are needed. Here, we display only the results on the low dimension dataset (Table 3).
\begin{table}[ht]
\begin{center}
\caption{\itshape Canonical correlation results on real data. $d$ is the value of $d$ assumed for the true correlation matrix. We report the 1st and 2nd canonical correlations. }
\label{s:table3}
\begin{tabular}{cccc cccc ccc}
  \toprule
  \toprule
 &&\multicolumn{2}{c}{d=10} && 
				\multicolumn{2}{c}{d=5} && 	\multicolumn{2}{c}{d=2} \\ \cline{3-4} \cline{6-7} \cline{9-10}
    & &{1st} & {2nd}  &&  {1st} &  {2nd} && {1st} &{2nd}\\ \midrule
            PSCCA& &	0.9759&	0.9229 &&  0.8732& 0.8245 && 0.7979 & 0.6791	\\
            \\
$\text{PCAN}^*$& &	0.8884&	0.8528 && 0.8343	& 0.7733 && 0.8188 &0.7179 \\
\\
PMD& & 0.5858&	0.5368 &&0.5858 &0.5368 && 0.5858 & 0.5147	\\
\\
BCCA && 0.2624& 0.2550  && 0.1981 & 0.1786 && 0.1608& 0.0069 \\
		    \bottomrule
\end{tabular}
\end{center}
\end{table}
\section{\textbf{Discussion and Future Work}}

We proposed a probabilistic approach of correlation and canonical correlation analysis for sparse count data. PSCCA is a model-based approach to estimate correlations and canonical correlations at the natural parameter level rather than at the raw data level. Both the simulation study results and the real data application indicate that PSCCA compares favorably to existing methods.

We provided a theoretical justification to prove that correlation coefficients and canonical correlation coefficients calculated from the raw count data {\bf X} and {\bf Y} will be smaller in magnitude than the correlation coefficients and canonical correlation coefficients calculated from the natural parameters ${\bold{ \theta}}$ and ${\bold {\lambda}}$ in section 3. And this explains why PSCCA achieves more extreme correlation and canonical correlation estimations in real data application. Meanwhile, we demonstrate that horseshoe prior can handle the sparsity very well and this, probably, is due to the horseshoe prior not regularizing the parameters far from zero, which is very important in extracting the important variables that only strongly identified in the NGS data.

As the demand increases for integrative high-dimensional complex data analysis, PSCCA is a linear method which may not be appropriate for fitting the complicated nonlinear situations. Recently, researchers in computer science and engineering developed deep CCA \citep{Benton19} to extract the nonlinear associations and extended it to multiple views. However, deep CCA benefits from the expressive power of deep neural networks, which have a black box drawback in that they are not easy to interpret and understand. PSCCA is a model-based approach which estimates the dependency within and between two datasets as a joint task, thus, PSCCA is more interpretable. One direction for our work is to develop a probabilistic deep CCA to handle more complex data structures and provide interpretable results. Another possible extension of our model is to build a longitudinal framework to extract the dependency relationship between two or multiple co-occurring time series data.


\backmatter


\normalsize
\section*{Acknowledgements}

 The authors thank the discussions with Dr.Roger S.Zoh about PCAN during the early stage of the PSCCA study design and anonymous referee for very useful comments that improved the presentation of the paper.\vspace*{-8pt}


\normalsize
\vspace*{-8pt}

\bibliographystyle{biom} \bibliography{pscca}

\vspace*{-12pt}



\normalsize

\section{}

\label{lastpage}

\end{document}